\documentclass[conference]{IEEEtran}
\IEEEoverridecommandlockouts

\usepackage{array}
\usepackage{color}
\usepackage{tabularray}
\usepackage{balance}

\usepackage{cite}
\usepackage{amsmath,amssymb,amsfonts}
\usepackage{algorithmic}
\usepackage{graphicx}
\usepackage{textcomp}
\usepackage[table,xcdraw]{xcolor}
\newcommand{\ignore}[1]{ }
\def\BibTeX{{\rm B\kern-.05em{\sc i\kern-.025em b}\kern-.08em
    T\kern-.1667em\lower.7ex\hbox{E}\kern-.125emX}}

\usepackage{multirow}

\usepackage[hidelinks]{hyperref}

\usepackage{courier}
\usepackage{soul}
\usepackage{ragged2e}
\usepackage{bbding}
\usepackage{pifont}

\usepackage[left=0.59in, right=0.59in, top=0.60in, bottom=0.60in]{geometry}

\begin{document}
\title{
{\normalsize Late Breaking Results:}\\
Fortifying Neural Networks: Safeguarding Against Adversarial Attacks with Stochastic Computing



}

\makeatletter
\newcommand{\linebreakand}{%
  \end{@IEEEauthorhalign}
  \hfill\mbox{}\par
  \mbox{}\hfill\begin{@IEEEauthorhalign}
}
\makeatother

\author{
\IEEEauthorblockN{Faeze S. Banitaba}
\IEEEauthorblockA{
\textit{University of Louisiana at Lafayette}\\
Lafayette, LA, USA \\
faeze.banitaba@louisiana.edu}
\and
\IEEEauthorblockN{Sercan Aygun}
\IEEEauthorblockA{
\textit{University of Louisiana at Lafayette}\\
Lafayette, LA, USA \\
sercan.aygun@louisiana.edu}
\and
\IEEEauthorblockN{M. Hassan Najafi}
\IEEEauthorblockA{
\textit{Case Western Reverse University}\\
Cleveland, OH, USA \\
mhassan.najafi@case.edu}
}

\maketitle

\begin{abstract}
In neural network (NN) security, safeguarding model integrity and resilience against adversarial attacks has become paramount.\ignore{
In the realm of neural network (NN) security, safeguarding the integrity and resilience of models 
against adversarial attacks has become paramount.}This study investigates the application of stochastic computing (SC) as a novel mechanism to fortify NN models. 
The primary objective is to assess the efficacy of SC to mitigate the deleterious impact of attacks on NN results. Through a series of rigorous experiments and evaluations, we explore the resilience of NNs employing SC when subjected to adversarial attacks. Our findings reveal that SC introduces a robust layer of defense, significantly reducing the susceptibility of 
networks to attack-induced alterations in their outcomes. This research contributes novel 
insights into the development of more secure and reliable NN systems, essential for applications in sensitive domains where data integrity is of utmost concern.
\end{abstract}


\section{Introduction}
The widespread adoption of neural networks (NNs) has revolutionized various sectors by providing unparalleled data processing, pattern recognition, and decision-making capabilities. Yet, this integration into crucial areas underscores a pressing issue: safeguarding these systems from increasingly complex adversarial attacks that aim to distort NN outputs, a critical concern in applications demanding high accuracy and reliability. Traditional security measures often fall short against these advanced threats, sparking interest in innovative defenses. Stochastic computing (SC) emerges as a notable strategy offering high error tolerance due to its inherent randomness, potentially increasing NN security against such manipulations. This study observes SC as a defense mechanism, assessing its potential to protect NN 
from adversarial attacks. 
The outcome 
is 
vital 
for sensitive sectors where 
accuracy and integrity are critical, aiming to foster the development of more 
robust systems.

In essence, this work 
offers insights into the potential of SC to enhance 
NN security, marking a step forward in the ongoing quest to develop robust 
artificial intelligence (AI) systems.
We propose 
using stochastic 
computations as a defense mechanism to make the network robust against attack. We will show how effective this mechanism is and 
study the impact of implementing each layer in the stochastic domain. 
We observe 
that implementing the first layers with SC can significantly improve the robustness of the network.
The key advantage of our proposed approach is that it eliminates the system's need to detect or block the attacker's data manipulations. Instead, our approach effectively sidesteps their threats, ensuring even in the face of successful attacks, we can 
deliver accurate outcomes.
Moreover, our method's adaptability for both simple and complex networks makes it a versatile defense strategy, suitable for any network facing the threat of adversarial attacks.

\section{Deep Learning and Adversarial Attacks}
NNs are frequently employed in systems where safety is paramount.
Classification NNs are a subset of machine learning models designed to 
classify input data into some predefined classes. 
These networks, built upon layers of interconnected nodes, 
learn to classify 
from training data. The versatility of classification NNs makes them popular 
for various applications, 
from image processing and speech recognition to medical diagnosis and financial analysis. 
While these 
NNs are powerful 
for data analysis and interpretation, they are vulnerable to attacks. 
The attacks can subtly manipulate input data, causing the network to misclassify. 
For example, an image slightly altered at the pixel level could be wrongly identified, or a text with minor alterations might be misinterpreted. This vulnerability poses significant challenges in applications where accuracy and reliability are crucial, such as in autonomous vehicles, security systems, and healthcare diagnostics. 

The emergence of \textit{adversarial attacks} 
has sparked a continuous interplay between the development of sophisticated attack techniques and the implementation of robust defenses~\cite{10427453}. This 
has fostered the creation of a diverse array of attack strategies, each tailored to challenge NNs under various scenarios. In response, the field has seen a parallel evolution in defensive measures, leading to a dynamic and ongoing advancement in securing 
NNs. There is a high demand for 
robust defense mechanisms to ensure the integrity and reliability of NNs. 

Within the dynamic landscape of machine learning and deep learning, the presence of adversarial attacks introduces major 
challenges. 
Deep learning models 
are particularly vulnerable to adversarial attacks.
These attacks involve slight modifications to inputs that, while imperceptible to humans, lead to incorrect classifications by the system. Examining such adversarial tactics, alongside countermeasures 
in deep learning 
has received considerable attention~\cite{thangaraju2022exploring}.
These strategic manipulations of 
data 
underscore the need for innovative defense strategies. 
The exploration of defense mechanisms extends beyond the adversarial spectrum, encompassing the compromise and exploitation of machine learning models and systems. In this intricate space, SC emerges as a potential safeguard, offering high resilience. 

SC is a computational model 
that represents and processes data in the form of random 
bit-streams rather than traditional binary formats. 
SC encodes values as the probability of observing 
a `$1$' in a random sequence of `1's and `0's. 
SC is known for its high fault tolerance and cost efficiency, making it particularly useful for applications such as signal processing and machine learning. 
SC systems offer a unique advantage in dealing with noisy or incomplete data and in scenarios where precision can be traded for 
lower power consumption and simplicity of hardware~\cite{10.1145/3611315.3633265}.
Earlier findings hinted at the SC design of 
convolutional NNs being partially resistant to adversarial attacks. Further studies introduced a hardware component known as the stochastic MAC-activation-pooling comparator. Implemented in the first layer, this unit can effectively counter the efforts of attackers~\cite{neugebauer2023stochastic}.

\section{SC Against Adversarial Attacks}

\subsection{Proposed Architecture}
In NNs, the first convolutional layers 
contribute significantly to the total area and power consumption due to the \textit{row}-\textit{column} nature of the process. The fundamental operation within these layers is 
dot-product, 
which multiplies 
the input data by the weights and then accumulates 
the results. This is typically followed by an activation function to introduce non-linearity and help the network learn complex patterns.

By integrating SC in inference, 
we can significantly enhance the resilience of the network to adversarial attacks. 
We perform multiplication operations, particularly in the first convolutional layers in the stochastic domain~\cite{faraji2019energy}.
This can be achieved by simple bit-wise \texttt{AND} operations, 
avoiding the complexity and cost associated with traditional 
binary multipliers. We exploit state-of-the-art quasi-random bit-streams for accurate multiplication with short bit-streams.
Our approach simplifies the computation process, enhancing its efficiency and saving 
resources. More importantly, the computations and, hence, NN become highly robust against error. 
Our solution maintains the accuracy of NNs even when they are under attack, such as from adversarial threats, thereby ensuring 
the system's security.

Stochastic processing 
involves encoding data to random bit-streams, 
where the value is depicted by the likelihood of observing 
a `1' in the sequence. Prior works showed that the savings from the simple computation logic (e.g., \texttt{AND} gate for multiplication) could well compensate for the overhead cost of converting data to bit-stream format. 
Our approach leaves the training phase of the network untouched and only replaces the multiplication operations in the inference phase. 
By simply modifying the convolution unit 
in the chosen layers, we enhance the network's robustness during the testing stage without revisiting the initial training process. This strategy ensures that existing NNs can be easily adapted to be more resilient against attacks without requiring comprehensive retraining or significant architectural alterations. 
The pivotal aspect of the proposed solution is in addressing 
the challenging task of reverting previously applied attacks. Even in the presence of manipulated data, employing stochastic computations 
alleviates the impact of malicious intentions.

\begin{figure}
\centering
	\includegraphics[trim={3 10 0 10},clip,width=\linewidth]{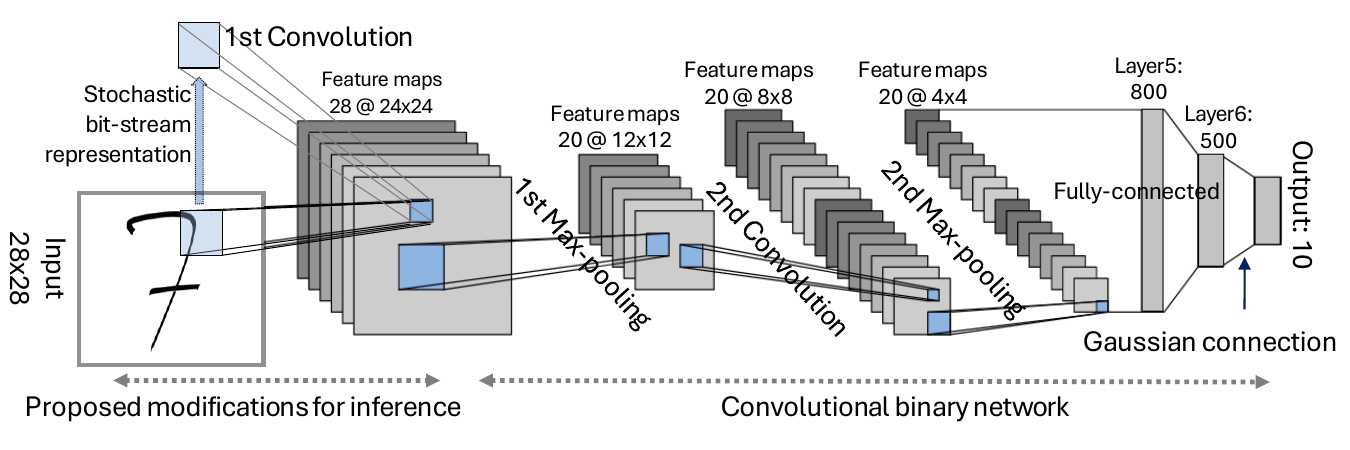}
\caption{Proposed robust NN setup for LeNet-5 model; the first convolution layer is equipped with our model, transitioning data 
 to the stochastic domain, followed by SC multiplications. 
}
	\label{lenet5}
\end{figure}

\subsection{Implementation}

We 
explore how SC can augment the network's ability to tolerate adversarial manipulations. 
First, we apply the proposed defense mechanism 
to the LeNet-5 network 
and then to 
ResNet-20 
as a more complex model.

Fig.~\ref{lenet5} shows the LeNet-5 model, which is used with the MNIST dataset for image classification. 
This model comprises five layers, including two 
convolution 
and three dense layers. After each 
layer, an activation function is applied. The first two utilize the rectified linear unit (ReLU) activation, while the final layer employs the `Softmax' function for output normalization. 
We convert the data 
from binary to quasi-random bit-streams in the targeted layer, varying the bit-stream lengths as needed. The multiplication operations are performed in the target layer by 
bit-wise \texttt{AND} 
between the bit-streams. For bit-stream generation, we employ Sobol sequences as described in~\cite{Sobol_TVLSI_2018}. 
All other operations 
within the network are conducted as per standard procedures.
We also extend our implementation to a more complex model, ResNet-20, classifying the CIFAR-10 dataset.  
As 
shown 
later, our proposed architecture demonstrates the capability to \textit{thwart attacks} across all tested network configurations. 

To evaluate the robustness against attacks, 
we simulate 
the targeted L2 attack, also known as the Carlini and Wagner (C\&W) attack~\cite{carlini2017towards}, a method grounded in one of the most common 
distance metrics for crafting adversarial examples. 
The L2 distance metric quantifies the Euclidean (root-mean-square) distance between 
the original input ($x$) and a perturbed version of $x$ ($x'$) that is close in terms of input space but misclassified in a different class $t$. The L2-norm attack exploits this metric to identify and generate $x'$. In our experiment, 
we assume the adversary has full knowledge of the NN, including its architecture and parameters, emulating a white-box attack. 
The expectation is that a defense mechanism effective against the L2 attack will likely be capable of counteracting other adversarial strategies. Our evaluation results 
demonstrate that our proposed architectural modifications significantly enhance the network's resilience against attacks based on the L2 distance metric.

\subsection{Results}
 
We conducted a series of experiments to gauge the impact of employing SC under attack stress for the two targeted NN models. 
Both networks were deliberately compromised throughout the tests to achieve zero accuracy, simulating a successful attack. Any deviation from this state, leading to a restoration of accuracy to ideal values, shows the success of the defense mechanism 
in shielding against adversarial attacks. 
Table~\ref{tab:mnist-ln} presents the classification performance of LeNet-5 on the MNIST dataset before and after 
attack. Prior to the attack, the presence of SC 
did not 
affect the results significantly, achieving comparable accuracy to that of the conventional non-SC implementation (99\%). 
We 
assessed the involvement of SC in the first and second layers independently. We observed that employing SC in the second layer does not 
yield advantageous outcomes, particularly for smaller bit-stream sizes. 
After the attack, 
substantial improvement is observed in the reported classification numbers, with the accuracy reaching up to 79\%, compensating for the initial 0\% accuracy resulting from the attacker's success across the entire test set.
Following our evaluation of LeNet-5, we explore 
ResNet-20 model, a notably more intricate architecture. To the best of our knowledge, this is the first time in the literature that 
SC has been employed to enhance defense mechanisms against security challenges in a complex network architecture. Table~\ref{tab:cifar-rn} reports the 
SC's impact on this model. 
We note that we need longer bit-streams (e.g., $N$=256) to achieve conventional network (non-SC) accuracy 
with this model. As can be seen, 
SC proves to be effective to defend 
the attack across all bit-stream sizes, efficiently mitigating the effects of malicious alterations on the network. We observe restored accuracy levels up to 85\%, healing from the worst-case scenario of 0\% accuracy resulting from the attacker's full success on the non-SC model.
Compared to state-of-the-art, 
our solution demonstrates a misclassification rate of 10\% in a safeguarded version using SC in the presence of an attack. 
In contrast, the state-of-the-art approach~\cite{neugebauer2023stochastic} exhibits a 79.1\% misclassification rate for the same 
CIFAR-10 dataset. 


\begin{table}[t]
\centering

\caption{\textbf{LeNet-5} Classification Accuracy for \textbf{MNIST} Dataset, When SC Applied to 1st or 2nd Convolution Layers}
\label{tab:mnist-ln}
\begin{tabular}{|c|c|c|c|c|} 
\hline
\multirow{2}{*}{\begin{tabular}[c]{@{}c@{}}Bit-stream\\Length\\(N)\end{tabular}} & \multicolumn{2}{c|}{Accuracy Before Attacks } & \multicolumn{2}{c|}{Accuracy \textbf{After} Attacks} \\ 
\cline{2-5}
 & \begin{tabular}[c]{@{}c@{}}SC in the\\1st Layer\end{tabular} & \begin{tabular}[c]{@{}c@{}}SC in the\\2nd Layer\end{tabular} & \begin{tabular}[c]{@{}c@{}}SC in the\\1st Layer\end{tabular} & \begin{tabular}[c]{@{}c@{}}SC in the\\2nd Layer\end{tabular} \\ 
\hline
8 & 0.93 & 0.56 & 0.69 & 0.28 \\ 
\hline
16 & 0.96 & 0.79 & \textbf{0.79} & 0.36 \\ 
\hline
32 & 0.98 & 0.93 & 0.61 & 0.41 \\ 
\hline
64 & 0.98 & 0.97 & 0.57 & 0.39 \\ 
\hline
1024 & 0.99 & 0.98 & 0.5 & 0.33 \\
\hline
\end{tabular}
\end{table}


\begin{table}[t]
\centering
\caption{\textbf{ResNet-20} classification accuracy for \textbf{CIFAR-10} dataset, while applying stochastic bit-stream processing}
\label{tab:cifar-rn}
\resizebox{\linewidth}{!}{%
\begin{tabular}{|>{\centering\hspace{0pt}}m{0.21\linewidth}|>{\centering\hspace{0pt}}m{0.24\linewidth}|>{\centering\hspace{0pt}}m{0.24\linewidth}|>{\centering\arraybackslash\hspace{0pt}}m{0.24\linewidth}|} 
\hline
Layer with SC & Bit-stream Length (N) & Accuracy Before Attacks when SC is applied & Accuracy \textbf{After} Attacks when SC is applied \\ 
\hline
1st & 8 & 0.31 & 0.32 \\ 
\hline
1st & 16 & 0.53 & 0.65 \\ 
\hline
1st & 32 & 0.72 & 0.72 \\ 
\hline
1st & 64 & 0.88 & 0.83 \\ 
\hline
1st & 128 & 0.87 & 0.83 \\ 
\hline
1st & 256 & 0.90 & 0.85 \\ 
\hline
1st \& 2nd & 128 / 128 & 0.89 & 0.83 \\ 
\hline
1st \& 2nd \& 3rd & 128 / 128 / 32 & 0.87 & 0.81 \\
\hline
\end{tabular}
}
\end{table}

\balance
\section{Conclusion}
 
This work proposed SC as an effective defense mechanism to fortify neural networks against adversarial attacks. 
We observed that incorporating SC 
notably strengthens the network resilience. 
We highlight SC's capacity to craft neural networks that are both accurate and reliable amidst adversarial challenges, signifying a major leap forward in achieving secure AI systems.
With additional research and exploration, the proposed method has the potential to emerge as a novel approach in the ongoing effort to mitigate threats to machine learning models. 


\bibliographystyle{unsrt2authabbrvpp}
\bibliography{References, Hassan}

\begin{thebibliography}{1}

\bibitem{10427453}
N.~M. Grigorieva and S.~A. Petrenko.
\newblock Known adversarial attacks by 2023.
\newblock In {\em 2023 Seminar on Information Systems Theory and Practice (ISTP)}, pp. 32--36, 2023.

\bibitem{thangaraju2022exploring}
A.~Thangaraju and C.~Merkel.
\newblock Exploring adversarial attacks and defenses in deep learning.
\newblock In {\em IEEE CONECCT}, 2022.

\bibitem{10.1145/3611315.3633265}
M.~S. Moghadam \textit{et~al.}
\newblock Accurate and energy-efficient stochastic computing with van der corput sequences.
\newblock In {\em Proc. of the 18th ACM}, NANOARCH '23, 2024.

\bibitem{neugebauer2023stochastic}
F.~Neugebauer \textit{et~al.}
\newblock Stochastic computing as a defence against adversarial attacks.
\newblock In {\em IFIP DSN-W}, 2023.

\bibitem{faraji2019energy}
S.~R. Faraji \textit{et~al.}
\newblock Energy-efficient convolutional neural networks with deterministic bit-stream processing.
\newblock In {\em DATE}, 2019.

\bibitem{Sobol_TVLSI_2018}
S.~Liu and J.~Han.
\newblock Toward energy-efficient stochastic circuits using parallel sobol sequences.
\newblock {\em IEEE Trans. Very Large Scale Integr. Syst.}, 26(7), 2018.

\bibitem{carlini2017towards}
N.~Carlini and D.~Wagner.
\newblock Towards evaluating the robustness of neural networks.
\newblock In {\em IEEE SP}, 2017.

\end{thebibliography}

\end{document}